# "From what I see, this makes sense:"
# Seeing meaning in algorithmic results

**Samir Passi**
Cornell University
Ithaca, NY, USA
sp966@cornell.edu

**Phoebe Sengers**
Cornell University
Ithaca, NY, USA
sengers@cornell.edu

**ABSTRACT**
In this workshop paper, we use an empirical example from our ongoing fieldwork, to showcase the complexity and situatedness of the process of making sense of algorithmic results; i.e. how to evaluate, validate, and contextualize algorithmic outputs. So far, in our research work, we have focused on such sense-making processes in data analytic learning environments such as classrooms and training workshops. Multiple moments in our fieldwork suggest that meaning, in data analytics, is constructed through an iterative and reflexive dialogue between data, code, assumptions, prior knowledge, and algorithmic results. A data analytic result is nothing short of a sociotechnical accomplishment – one in which it is extremely difficult, if not at times impossible, to clearly distinguish between 'human' and 'technical' forms of data analytic work. We conclude this paper with a set of questions that we would like to explore further in this workshop.

**Author Keywords**
Algorithms; Data Analysis; Learning; Sense-making

**ACM Classification Keywords**
H.m. [Information Systems]: Miscellaneous

**INTRODUCTION**
Algorithmic data analysis is enabling new ways to produce and validate knowledge [5, 6], generating interest in in HCI and social sciences in the roles and implications of algorithmic data analyses [see inter alia 1, 7**,** 9, 10]. We now know that algorithms can be selective [9], subjective [4], and biased [see 8 for a review]; that they work on multiple assumptions about the world and how it functions [2, 5, 9, 10]; and that they enable, facilitate, and constrain human action and knowledge [2, 3]. Algorithmic knowledge production is a deeply social practice with sociocultural, economic, and political groundings and consequences.

In our research, we focus on algorithmic knowledge production as a sociotechnical process, studying how people *(in our case: computer science, information science, and humanities students and scholars)* make sense of the world with and through algorithms. We study how people interpret, organize, and analyze the world through computational data forms, data analytic algorithms, and statistical techniques. While data analysis is often understood as the work of faceless, nameless, and unbiased numbers and algorithms, we are focusing on the situated, iterative, and discretionary human work required to see, represent, and organize the world algorithmically. Using ethnographic research methods, we explore questions such as: What forms of human and technical work are involved in data analyses? How do people evaluate and contextualize data analytic results to make them meaningful in social contexts? How do we know whether data analysis has 'worked'? How can we design technologies or modify work and organizational processes to better represent the allocation of data analytic labor across humans and machines?

We are currently studying such questions in data analytic learning environments such as classes and training workshops. Of course, classroom students and workshop participants are not representative of all data analysts; we acknowledge that working with algorithms in a classroom or workshop is different in nature from working with them in a company or research center. What learning environments do provide is a context within which certain aspects of a practice become particularly apparent, through the explicit focus on describing, explaining, and demonstrating established methods, tools, and techniques. This helps us not only better understand the underlying assumptions in seemingly basic, yet foundational, aspects of data analysis but also observe how pedagogic demonstrations and analytical examples enable or facilitate specific data analytic norms and heuristics.

So far, we have conducted three sets of ethnographic fieldwork at a major U.S. East Coast university. *First*, we conducted 4 months of participant-observation in a senior/graduate level machine learning class in fall 2014. One of the authors was enrolled as a student in a section of the course with ~80 students. The students in this class mainly consisted of computer and information science majors and graduate students. *Second*, we conducted participant-observation in a series of three digital humanities workshops organized by a different professor at the same university during spring 2015. Broadly put, in the digital humanities, humanists and computer and information scientists use both computational and interpretive methods to analyze data such as that in history, literature, etc. The purpose of the workshops was to expose students to computational techniques for text analyses. The majority of the participants were doctoral students in the English department. *Third*, we conducted 4 months of participant-observation in a senior/graduate level text mining and topic modeling course taught in fall 2015. One of the authors was enrolled as a student in the course with ~30 students. The students in this course consisted of computer and information science majors, and graduate humanities students.

A finding from our research that is particularly relevant to this workshop is about the complexity and situatedness of how classroom students and workshop participants make sense of algorithmic results; i.e. how they evaluate, validate, and contextualize algorithmic outputs. The process of making sense of algorithmic results is neither linear nor straightforward. Applying an algorithm involves a series of situated decisions to iteratively, often creatively, adapt prior knowledge, data analytic algorithms, and contingent empirical data to each other. Multiple moments in our three sets of fieldwork suggest that meaning, in data analytics, is constructed through an iterative and reflexive dialogue between data, code, assumptions, prior knowledge, and algorithmic results. A data analytic result is nothing short of a sociotechnical accomplishment – one in which it is extremely difficult, if not at times impossible, to clearly distinguish between 'human' and 'technical' forms of data analytic work. Below we provide a small example to illustrate an instance of such sense making.

**EMPIRICAL EXAMPLE**

This example is from the text mining and topic modeling class. The instructor of this class has been extensively involved in developing computational methodologies for humanities and social sciences, and has been collaborating with humanists for a long period of time. His academic experiences include undergraduate work in Greek, Latin, and computer science, and graduate work in statistical data mining. In his own work, he develops and researches the use of data mining tools and methods for studying history and literature.

At the point we pick up this example, the instructor has already introduced the concept of 'topic models' to the students. A topic model is a mathematical view of 'topics' in a set of documents. One can think of these topics as discursive themes. A seemingly simple, yet foundational, way of doing this algorithmically is to keep track of what words occur together more frequently than what would be expected at random. The intuition behind this is that a topic has an associated semantic universe, and thus certain groups of words are more likely to occur together when they are used to refer to a particular topic/theme.

In this class, instructor and students are analyzing a set of Danish folk tales through an existing topic-modeling algorithm. They are using a web-based implementation of this algorithm that provides a list of, what the algorithm believes to be, topics. A topic is presented as a collection of words (in decreasing order of relevance to the topic). Clicking on a topic provides snippets of underlying text that, according to the algorithm, are characteristic of the selected topic. Figure 1 shows a screenshot of what the algorithm's output looks like for the Danish folk-tales.

Once the algorithm is done analyzing the set of Danish folk-tales, the instructor scrolls through the outputted topics, choosing one that catches his attention (figure 1).

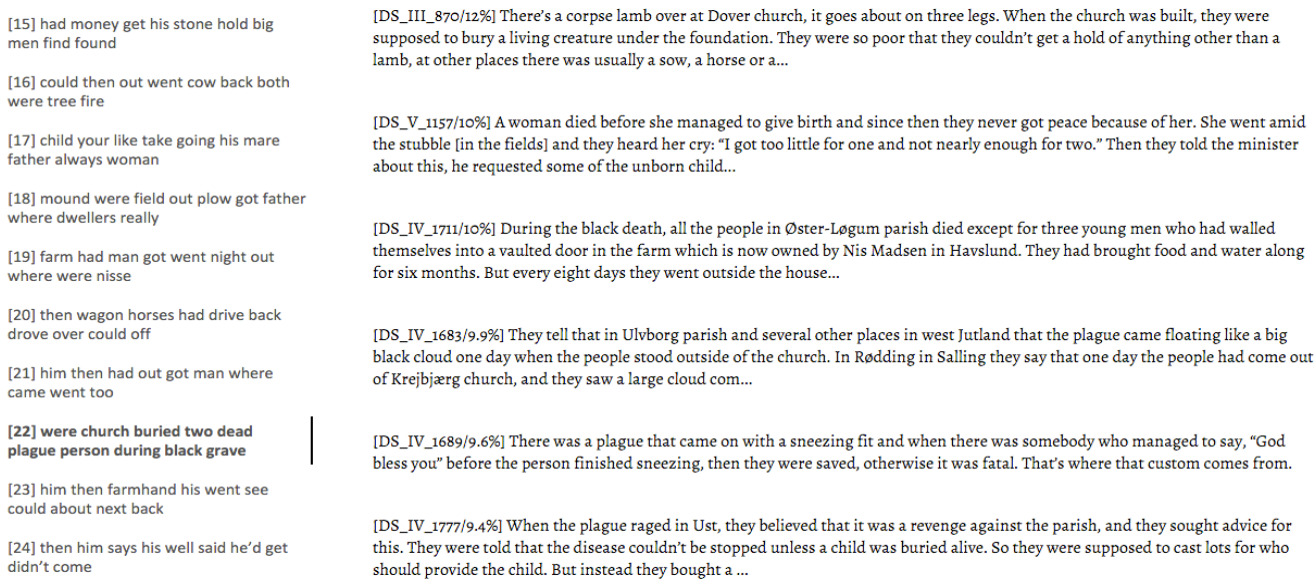

Figure 1. Algorithmic result of analyzing a set of Danish folk-tales

The words in this topic are: *were church buried dead plague person during black grave.* The instructor clicks on the topic to bring up the text snippets, gazes at the snippets for a while, looks back at the students, and asks them: "does this topic make sense? Do these snippets on the right mean anything to us?" The students remain quiet, perhaps pondering over the relation between the topic and the snippets. After a while, some of the students start to nod. At this point the instructor describes his own thought process of how he made sense of what he saw:

> *We are seeing a lot of 'plague,' but it is interesting that this word is linked to the word 'church.' This is one of the reasons that folklorists might find this interesting. These stories [i.e., the Danish folk tales] are individually not that interesting, but together there might be a pattern that might be interesting to look at. For instance, is there a relation between religious figures and plague? We don't get an answer here, but we get a clue that there might be something here.*

The instructor then moves on to select another topic. The words in this topic are: *mound were field out plow got father where dwellers really.* The instructor again stares at the topic and snippets for a while, and then asks the students: "what do we think of this one? Does it make sense too?" More silence ensues. Seeing that there aren't any nods of approval this time around, the instructor explains:

> *Just seeing this group of words, at first sight, might not seem to mean anything. It is at this point that knowing something about the underlying data helps. In this case, for instance, we are seeing this topic because there is a known category of characters in Danish folktales called 'mound dwellers.' How many of you knew before this that there was this category in Danish folk-tales? [No one raises his or her hands.] Well, now you know it. The topic model told you so.*

Finally, the instructor chooses a third topic with the following words: *mill road often robbers woods pond role through ago across.* But, instead of asking the same question, this time he phrases his question differently:

> *So, this topic here. There is a mill. There is a road. There are robbers on the road, and a rope involved. Can we tell a story about this? Is this a topic? From what I see, this makes sense.*

Most, if not all, students start to nod. Robbers in the woods, on a road, with a rope – that's got to make sense, right?

**DISCUSSION**

Even in this small example, we are clearly able to see how the classroom instructor straddles between data, code, prior knowledge, and on-screen results to teach students ways to make sense of what the algorithm is saying. The results are never discussed as – 'here is a topic.' Instead, a set of question is posed: What do we think of this output? Does it make sense? Can we tell a story using this? These questions in themselves signify the process of *making* sense of the results instead of simply *accepting* them as truth claims. What is also important is the role prior knowledge and expectations play in making sense of such results. Does a result conform to what we already know or believe in (e.g., robbers in the woods)? Or, does the result tell us something we did not already know (e.g., Danish mound dwellers)? Indeed, at times (though not in this example) we saw how classroom students and workshop participants struggled to make sense of certain results when they did not conform to their existing worldviews and beliefs. The aforementioned example was of topic modeling: a specific type of data analysis – one in which the output is a combination of words (as parts of topics) and numbers (as a series of probabilities for words in topics). However, we observed similar sense-making processes in our research work in other forms of data analysis as well such as similarity clustering, classification, ensemble learning, etc.

Moreover, such sense making is not restricted to results. In our research, we observe forms of meaning making – through choices, decisions, and interpretations – in almost every aspect of data analytics. What forms of questions can be answered through data analysis? What data do we need to answer a given question? How much data is required to answer a question? What parts of the data need to be cleaned, and why? How do we treat outliers? These, and many more, questions form the core of data analysis as observed in classrooms and workshops.

The questions we would like to explore through this workshop range from how we can study and describe such forms of data analytic sense making to how can we modify existing processes, or create new ones, to better depict the multiple forms of choices, decisions, and interpretations that go into every data analytic project. Working on questions such as these, we believe, can help us contextualize algorithmic knowledge as well as make us better understand and communicate the limits of such knowledge. Moreover, providing an explicit description of the situated decisions that data analytic researchers take can not only help people get a more nuanced understanding of technical choices and algorithmic results but also help data analysts think through aspects of their work that although may seem 'non-technical,' greatly impact their results and findings. Algorithms don't produce knowledge on their own. We produce knowledge *with* and *through* algorithms and datasets. A human centered view of data analytic work can help put the human back in the algorithm, providing a more accurate description of the division of algorithmic labor.

**AUTHOR BIOS**
Samir Passi joined Cornell University in 2013 as a Ph.D. student in Information Science at Cornell University to work with Phoebe Sengers, Steve Jackson, and Malte

Ziewitz. His research interests lie primarily at the intersection of Information Science and Science & Technology Studies. In his work – situated within social studies of data analysis – he looks at how people work with and through machines to make sense of data. Samir has a research masters in Science & Technology Studies from Maastricht University, The Netherlands, and an undergraduate degree in Information & Communication Technology engineering from DA-IICT, India.

Phoebe Sengers is a faculty member in Information Science and Science & Technology Studies at Cornell University, where she leads the Culturally Embedded Computing group. Her work integrates analysis of the political and social impacts of technology with IT design. Her primary current focus is a long-term design-ethnographic and historical study of sociotechnical change in the small, traditional fishing community of Change Islands, Newfoundland, looking at how changing sociotechnical infrastructures are tied with changing orientations to time, technology, and labor. Phoebe graduated from Carnegie Mellon University in 1998 with a self-defined interdisciplinary Ph.D. in Artificial Intelligence and Cultural Theory.

**ACKNOWLEDGMENTS**

This research is supported under U.S. National Science Foundation Grant #1526155.

**REFERENCES**
1. Tom Boellstorff. 2013. Making big data, in theory. *First Monday,* 18, 10.
2. Geoffrey C. Bowker. 2013. Data Flakes: An Afterword to "Raw Data" Is an Oxymoron. In *Raw Data Is an Oxymoron,* Lisa Gitelman (ed.). MIT Press, Cambridge, 167-171.
3. Geoffrey C. Bowker. 2014. The Theory/Data Thing. *International Journal of Communication,* 8: 1795-1799.
4. danah boyd and Kate Crawford. 2012. Critical Questions for Big Data: Provocations for a cultural, technological, and scholarly phenomenon. *Information, Communication & Society,* 15, 5: 662-679.
5. Lisa Gitelman. 2006. *Raw Data Is an Oxymoron.* MIT Press.
6. Sabina Leonelli. 2014. What differences does quantity make? On the epistemology of Big Data in biology. *Big Data & Society,* 1, 1, 1-11.
7. Kathleen H. Pine and Max Liboiron. 2015. The Politics of Measurement and Action. In *Proceedings of the SIGCHI Conference on Human Factors in Computing Systems* (CHI '15), 3147-3156.
8. Solon Barocas. Data Mining and the Discourse on Discrimination. In *Proceedings of Data Ethics Workshop at ACM Conference on Knowledge Discovery and Data-Mining (KDD),* 1-4.
9. Tarleton Gillespie. 2014. The Relevance of Algorithms. In *Media Technologies: Essays on Communication, Materiality, and Society,* Tarleton Gillespie, Pablo J. Boczkowski and Kirsten A. Foot (eds.). MIT Press, Cambridge, 167-194.
10. Alex S. Taylor, Siân Lindley, Tim Regan, David Sweeney, Vasillis Vlachokyriakos, Lille Grainger, and Jessica Lingel. 2015. Data-in-place: Thinking through the Relations Between Data and Community. In *Proceedings of the SIGCHI Conference on Human Factors in Computing Systems* (CHI '15), 2863-2872.